\documentclass[%
 reprint,
 amsmath,amssymb,
 aps,
 pre,
]{revtex4-2}

\usepackage{graphicx}
\usepackage{dcolumn}
\usepackage{bm}
\usepackage{xcolor}

\begin{document}

\preprint{APS/123-QED}

\title{Macroscopic Dynamics of Neural Networks with Heterogeneous Spiking Thresholds}

\author{Richard Gast}
 \email{richard.gast@northwestern.edu}
\author{Sara A. Solla}%
\author{Ann Kennedy}%
\affiliation{%
 Department of Neuroscience, Feinberg School of Medicine, Northwestern University, Chicago, USA.\\
}%



\date{\today}

\begin{abstract}
Mean-field theory links the physiological properties of individual neurons to the emergent dynamics of neural population activity. 
These models provide an essential tool for studying brain function at different scales; however, for their application to neural populations on large scale, they need to account for differences between distinct neuron types.
The Izhikevich single neuron model can account for a broad range of different neuron types and spiking patterns, thus rendering it an optimal candidate for a mean-field theoretic treatment of brain dynamics in heterogeneous networks.
Here, we derive the mean-field equations for networks of all-to-all coupled Izhikevich neurons with heterogeneous spiking thresholds.
Using methods from bifurcation theory, we examine the conditions under which the mean-field theory accurately predicts the dynamics of the Izhikevich neuron network.
To this end, we focus on three important features of the Izhikevich model that are subject here to simplifying assumptions: (i) spike-frequency adaptation, (ii) the spike reset conditions, and (iii) the distribution of single-cell spike thresholds across neurons.

Our results indicate that, while the mean-field model is not an exact model of the Izhikevich network dynamics, it faithfully captures its different dynamic regimes and phase transitions. 
We thus present a mean-field model that can represent different neuron types and spiking dynamics.
The model is comprised of biophysical state variables and parameters, incorporates realistic spike resetting conditions, and accounts for heterogeneity in neural spiking thresholds.
These features allow for a broad applicability of the model as well as for a direct comparison to experimental data.
\end{abstract}

\maketitle


\section{\label{sec:intro}Mean-field dynamics of populations with different neuron types}

Mathematical models are a necessary tool for understanding brain function and dynamics \cite{dayan_theoretical_2001,izhikevich_dynamical_2007}.
Due to the vast number of neurons and synapses in the brain, methods from statistical physics and mean-field theory provide a powerful tool for modeling its mesoscale dynamics \cite{deco_dynamic_2008,coombes_large-scale_2010,chialvo_emergent_2010}.
Classic mean-field models apply heuristic arguments derived from experimental data to propose equations that govern the evolution of averaged quantities such as population firing rates or mean post-synaptic potentials  \cite{wilson_excitatory_1972, lopes_da_silva_model_1974, jansen_electroencephalogram_1995, robinson_propagation_1997}.
While these classical models have contributed to our understanding of interacting neural populations within and across brain areas, they do not account for phenomena emerging from spike synchronization, nor do they relate single-cell properties to mean-field dynamics \cite{devalle_firing_2017,coombes_next_2019}. 
A more recent formulation of mean-field theory derives a set of closed-form mean-field equations from the evolution equations of a set of all-to-all coupled spiking neurons, therefore overcoming these problems \cite{luke_complete_2013,montbrio_macroscopic_2015,bick_understanding_2020}.

Mean-field equations derived from spiking neurons enable the study of the effects of heterogeneously distributed single cell parameters at the mean-field level.
Unfortunately, the spiking neural networks for which mean-field equations have been derived so far are defined based on dimensionless state variables, such as the phase on the unit circle or a dimensionless representation of a membrane potential \cite{luke_complete_2013,montbrio_macroscopic_2015}.
Here, we apply this approach to the derivation of mean-field equations for networks of coupled Izhikevich (IK) neurons, which comes in two different versions: a dimensionless version and a version with state variables with physical units \cite{izhikevich_simple_2003,izhikevich_dynamical_2007}.
Using the latter version of the IK model, the parameters of the IK neurons can be specified through recordings of individual cell properties such as cell membrane capacitance, resting membrane potential, or firing threshold \cite{izhikevich_dynamical_2007}. 
Furthermore, the IK neuron model can represent a wide range of neuron types and neural firing patterns, thus providing an ideal model for large-scale models of the  dynamics of heterogeneous neural populations \cite{izhikevich_simple_2003,izhikevich_which_2004,izhikevich_large-scale_2008}.

Deriving the mean-field equations for networks of IK neurons represents a challenge.
The essence of the mean-field theory that has been successfully applied to quadratic integrate-and-fire neurons and theta neurons \cite{luke_complete_2013,montbrio_macroscopic_2015} lies in the ansatz that the state variables of all neurons in the population are fully captured at all times by a Lorentzian probability distribution.
This ansatz, known as the Lorentzian ansatz, is mathematically equivalent to the Ott-Antonsen ansatz \cite{ott_low_2008,montbrio_macroscopic_2015}.
Crucially, the ansatz requires that the dynamics of a single neuron can be reduced to a single state variable: its phase on the unit circle.
Since the IK neuron model is a two-dimensional neuron model, this reduction is not possible without further simplifying assumptions.

In the remainder of this article, we derive the mean-field equations for networks of all-to-all coupled IK neurons with distributed firing thresholds and analyze how the underlying simplifying assumptions affect the mean-field dynamics of a network of IK neurons. 
We show that the mean-field model accurately captures a wide range of dynamic regimes and phase transitions of the underlying spiking network.
Furthermore, we analyze the conditions under which the mean-field predictions become less accurate.
These conditions include (a) strong spike-frequency adaptation at the single cell level, (b) narrow spike reset conditions, and (c) strong neural heterogeneity.
We relate these conditions to the simplifying assumptions used in the derivation of the mean-field equations, and show that even in these cases the mean-field predictions capture the qualitative properties of the bifurcation diagrams of the corresponding spiking networks, although the quantitative fit becomes worse.
Finally, we provide a correction term that accounts for narrow spike reset conditions.

\section{\label{sec:results1}Mean-Field Models of Coupled Izhikevich Neurons}

\subsection{The Spiking Neural Network}

We consider networks of coupled Izhikevich (IK) neurons of the form

\begin{align}
    C \dot v_i &= k (v_i - v_r)(v_i - v_{\theta,i}) - u_i + I + g s (E-v_i), \label{eq:v_i}\\
    \tau_u \dot u_i &= - u_i + b(v_i - v_r) + \tau_u \kappa \delta(v_i-v_p), \label{eq:u_i}
\end{align}

where $v_i$ and $u_i$ represent the membrane potential and the membrane recovery variable of the $i^{th}$ neuron in a network \cite{izhikevich_dynamical_2007}.
This neuron is defined to spike when $v_i \geq v_p$, where $v_p$ is the peak membrane potential; when this condition is met, a spike is counted and $v_i$ is reset to the reset potential $v_0$.
The recovery variable $u_i$ is driven by two terms.
The term $\kappa \delta(v_i-v_p)$ in the right-hand side of Eq.\eqref{eq:u_i}, where $\delta$ is the Dirac delta function,  represents an increase of $u_i$ by $\kappa$ whenever the neuron spikes.
This introduces a spike-frequency adaptation mechanism into the neuron model, since $u_i$ enters into Eq.\eqref{eq:v_i} as a hyperpolarizing variable.
The term $b(v_i-v_r)$ provides a "restoring force" that indirectly drives the membrane potential to its resting value $v_r$ in the absence of external input.
Additional parameters that control the behavior of the neuron are the cell capacitance $C$, the leakage parameter $k$, the spike threshold potential $v_{\theta}$, and the recovery variable time constant $\tau_u$.
Finally, the neuron in Eq.\eqref{eq:v_i} receives two forms of input current: an extrinsic current $I$, and a synaptic current that depends on the synaptic activation $s$, the maximum synaptic conductance $g$, and the reversal potential $E$.
We model the synaptic activation $s$ as the convolution of the mean-field activity of the network with an exponential activation kernel; this can be expressed as a first-order differential equation of the form

\begin{equation}
    \tau_s \dot s = -s + \frac{J \tau_s}{N} \sum_{j=1}^{N} \delta(v_j-v_p), \label{eq:s}
\end{equation}

where $\tau_s$ is a decay time constant and $J$ is a global coupling constant.  
Thus, Eq.\eqref{eq:s} represents the synaptic activation of each neuron in an all-to-all coupled network of $N$ neurons.

It has been shown that the population dynamics of certain families of spiking neural networks are fully captured by their average firing rate and average membrane potential, and that their mean-field equations can be derived via the Ott-Antonsen ansatz or the equivalent Lorentzian ansatz \cite{ott_low_2008,montbrio_macroscopic_2015,bick_understanding_2020}.
Most recently, a study has shown that the mean-field equations for a system of abstract, dimensionless IK neurons can be derived using a similar approximation \cite{chen_exact_2022}.
We will follow the latter approach to derive the mean-field equations for the heterogeneous spiking neural network given by eqs.(\ref{eq:v_i}-\ref{eq:s}). 

\subsection{Incorporating Neural Heterogeneity into Mean-field Models}

One important aspect of the spiking neural network considered here is that it allows for heterogeneity across neurons in the network.
Typically, dimensionless mean-field models incorporate spiking heterogeneity by treating the input variable $I = \eta_i + I_{ext}(t)$ as a distributed quantity, with neuron-specific background input $\eta_i$ and global extrinsic input $I_{ext}(t)$. 
The spike threshold $v_{\theta}$ has also been related to single cell heterogeneity \cite{wilson_excitatory_1972,rich_loss_2022}.

While the input $I$ enters Eq.\eqref{eq:v_i} as an isolated term, the threshold  $v_{\theta}$ is multiplied by the state variable $v_i$; a  distributions of values of $v_{\theta}$ in the population thus couples nonlinearly to the membrane potential dynamics of the neuron.
So, while distributions over $I$ can represent heterogeneity in the tonic drive to a population, distributions over $v_{\theta}$ represent heterogeneity of the electrophysiological properties across cells within a population.
Another important difference between these two sources of neural heterogeneity is their experimental accessibility.
Spike thresholds can be measured in single cells via patch-clamp recordings and slow input current ramps.
The form of the distributions of spike thresholds across cells can be chosen to capture the results of such recordings.
On the other hand, background current distributions are a lumped representation of all input currents to a cell that are not explicitly incorporated in the model, and are thus much harder to infer from neural recordings.

For these reasons, we focus on $v_{\theta}$ as the heterogeneity parameter.
The values of $v_{\theta,i}$ in the network model are assumed to be neuron specific and drawn from a probability distribution $p(v_{\theta})$.

\subsection{\label{sec:derivation}Derivation of the Mean-field Equations}

We consider the system given by eqs.(\ref{eq:v_i}-\ref{eq:s}) in the thermodynamic limit, i.e. when $N \rightarrow \infty$.
In this limit, the state of the system can be defined via a density function $\rho(v,u,v_{\theta},t)$.
For a given neuron, this quantity represents the joint probability density of its spike threshold $v_{\theta}$, membrane potential $v$, and recovery variable $u$ at time $t$.   
The conservation of the number of neurons implies that the probability density $\rho(v,u,v_{\theta},t)$ must satisfy a continuity equation

\begin{align}
    \frac{\partial}{\partial t} \rho(v,u,v_{\theta},t) &= - \frac{\partial}{\partial v} [\rho(v,u,v_{\theta},t) G(v,u,s,v_{\theta})], \label{eq:continuity} \\
    G(v,u,s,v_{\theta}) &= \begin{pmatrix}
    G^v(v,u,s,v_{\theta}) \\
    G^u(v,u)
    \end{pmatrix}, \label{eq:flux} \\
\end{align}

where the right-hand side of Eq.\eqref{eq:continuity} represents the probability flux given by the vector field $G$ defined as

\begin{align}
    G^v &= \frac{1}{C} [k (v - v_r)(v - v_{\theta}) - u + I + g s (E-v)], \label{eq:vecfield_v}\\
    G^u &= \frac{1}{\tau_u} [b(v - v_r) - u] + \kappa \delta(v-v_p). \label{eq:vecfield_u} 
\end{align}

The order parameters for which we wish to derive mean-field equations are the average firing rate $r(t)$, the average membrane potential $v(t)$, and the average recovery variable $u(t)$, where averages are evaluated across neurons.
These order parameters can be defined in terms of Eq.\eqref{eq:continuity} via the following integrals:

\begin{align}
    r(t) &= \int_{v_{\theta}} \int_u G^v(v_p,u,s,v_{\theta}) \rho(v_p,u,v_{\theta},t) d u\, d v_{\theta}, \label{eq:r}\\
    v(t) &= \int_{v_{\theta}} \int_u \int_v v \rho(v,u,v_{\theta},t) d v\, d u\, d v_{\theta}, \label{eq:v}\\
    u(t) &= \int_{v_{\theta}} \int_v \int_u u \rho(v,u,v_{\theta},t) d u\, d v\, d v_{\theta}, \label{eq:u}
\end{align}

While eqs.(\ref{eq:v}-\ref{eq:u}) are simply the expected values of $u$ and $v$, 
Eq.\eqref{eq:r} represents the probability flux at $v = v_p$ (that is, the proportion of neurons emitting a spike at time $t$) under the assumption that $v_p \rightarrow \infty$ and $v_0 \rightarrow -\infty$.
We evaluate Eq.\eqref{eq:u} by following the approach outlined in \cite{chen_exact_2022}, which critically assumes that $u \gg \kappa$ for any $v_{\theta}$ - that is, for any neuron in the population.
This regime amounts to assuming that spike-frequency adaptation in the model is small.
Under this assumption, the dynamics of $u(t)$ can be approximated by replacing $\delta(v_i-v_p)$ with the average firing rate across neurons $r(t) = \frac{1}{N} \sum_{j=1}^{N} \delta(v_j-v_p)$:

\begin{equation}
    \tau_u \dot{u} = b (v-v_r) - u + \tau_u \kappa r. \label{eq:du} 
\end{equation}

With this approximation, the continuity equation \eqref{eq:continuity} can be integrated with respect to $u$ to yield

\begin{equation}
    \frac{\partial}{\partial t} \rho(v,t|v_{\theta}) = - \frac{\partial}{\partial v} [G^v(v,u,s,v_{\theta}) \rho(v,t|v_{\theta})], \label{eq:cont_cond}
\end{equation}

where we additionally used that $\rho(v,v_{\theta},t) = \rho(v,t|v_{\theta}) p(v_{\theta})$.
For a more detailed description of the derivation outlined above, see \cite{chen_exact_2022}.

To obtain expressions for $r$ and $v$, we apply the Lorentzian ansatz outlined in \cite{montbrio_macroscopic_2015}.
We assume that the distribution over $v$ can be fully captured at any time $t$ by a Lorentzian probability distribution

\begin{equation}
    \rho(v,t|v_{\theta}) = \frac{1}{\pi} \frac{x(t,v_{\theta})}{[v-y(t,v_{\theta})]^2 + x(t,v_{\theta})^2}, \label{eq:rho_lorentzian}
\end{equation}

centered at $y$ and with half-width-at-half-maximum $x$.
As shown in \cite{montbrio_macroscopic_2015}, these two parameters of the Lorentzian distribution are inherently related to $r$ and $v$ via

\begin{align}
    \frac{C \pi}{k} r(t) &= \int_{v_{\theta}} x(t,v_{\theta}) p(v_{\theta}) d v_{\theta} = x(t), \label{eq:rx}\\
    v(t) &= \int_{v_{\theta}} y(t,v_{\theta}) p(v_{\theta}) d v_{\theta} = y(t). \label{eq:vy}
\end{align}

By plugging Eq.\eqref{eq:rho_lorentzian} into Eq.\eqref{eq:cont_cond} and equating the left- and right-hand-side in powers of $v$, we find that the system dynamics can be described by a single complex variable $z(t,v_{\theta}) = x(t,v_{\theta}) + iy(t,v_{\theta})$, the dynamics of which obey

\begin{equation}
    C \frac{\partial}{\partial t} z(t,v_{\theta}) = i[-kz(t,v_{\theta})^2 + i \alpha z(t,v_{\theta}) + \beta], \label{eq:z}
\end{equation}
where $\alpha$ and $\beta$ are defined as
\begin{align}
    \alpha &= k(v_r+v_{\theta}) + g s, \label{eq:alpha}\\
    \beta &= k v_r v_{\theta} + g s E - u + I. \label{eq:beta}
\end{align}

Finally, to derive the equations for $r(t)$ and $ v(t)$ from Eq.\eqref{eq:z}, we would like to solve the integral

\begin{equation}
    \dot z = \frac{1}{C} \int_{v_{\theta}} \frac{\partial}{\partial t} z(t,v_{\theta}) p(v_{\theta}) d v_{\theta}. \label{eq:z_dot}
\end{equation}

As shown in \cite{montbrio_macroscopic_2015}, this integral can be evaluated analytically if $p(v_{\theta})$, the distribution of the heterogeneous spike threshold, is chosen to be a Lorentzian density function

\begin{equation}
    p(v_{\theta}) = \frac{1}{\pi} \frac{\Delta_v}{[v_{\theta} - \bar v_{\theta}]^2 + \Delta_v^2}, \label{eq:p_lorentzian}
\end{equation}

centered at $\bar v_{\theta}$ and with half-width-at-half-maximum $\Delta_v$.
For this choice of $p(v_{\theta})$ we can solve Eq.\eqref{eq:z_dot} by evaluating $\frac{\partial}{\partial t} z(t,v_{\theta})$ at the single pole of the integrand in the upper half of the complex plane $v_{\theta} = \bar v_{\theta} + i \Delta_v$.
Under further consideration of Eq.\eqref{eq:rx} and Eq.\eqref{eq:vy}, it holds that $z(t,\bar v_{\theta} + i \Delta_v) = x(t) + i y(t) = \frac{\pi C}{k} r(t) + i v(t)$.
By plugging this relationship into Eq.\eqref{eq:z_dot} and solving for $r$ and $v$, we obtain the following set of coupled ordinary differential equations 

\begin{align}
    C \dot{r} = &\frac{\Delta_v k^2}{\pi C} (v-v_r) + r(k(2 v - v_r - \bar v_{\theta}) - g s), \label{eq:r_dot}\\
    C \dot{v} = &k v(v - v_r - \bar v_{\theta}) - \pi C r(\Delta_v + \frac{\pi C}{k} r) \label{eq:v_dot}\\
    &+ k v_r \bar v_{\theta} - u + I  + g s (E-v), \nonumber \\
    \tau_u \dot{u} = &b(v-v_r) - u + \tau_u \kappa r, \label{eq:u_dot}\\
    \tau_s \dot s = &-s + \tau_s J r. \label{eq:s_dot}
\end{align}

For the derivation of Eq.\eqref{eq:s_dot}, we used Eq.\eqref{eq:s} together with $r(t) = \frac{1}{N} \sum_{j=1}^{N} \delta(v_j-v_p)$.
Under the assumptions that spike-frequency adaptation is small and that spike peak and reset potentials approach positive and negative infinity, respectively, this final set of four coupled ordinary differential equations fully captures the macroscopic dynamics of the spiking neural network given by eqs.(\ref{eq:v_i}-\ref{eq:s}).
Below, we demonstrate via numerical comparisons of the dynamics of the mean-field model and spiking neural network model that this is indeed the case.
Furthermore, we analyze how well the mean-field predictions describe the macroscopic dynamics of the spiking neural network when each of the assumptions on which the mean-filed derivation is based is violated. 
Finally, we analyze the quality of mean-field predictions when the assumption of a Lorentzian distribution of the spike threshold heterogeneity is violated in the spiking neural network.
To this end, we truncate the heavy tails of the Lorentzian probability distribution at different spike thresholds and study its effect on the mean-field dynamics of the spiking neural network.
This modification on the assumed form of $p(v_{\theta})$ accounts for the biological fact that spike thresholds are confined to a finite range of potentials, bound by the resting membrane potential from below and the peak membrane potential from above.

\subsection{Form of the recovery variable $u$}

We derived the mean-field equations (\ref{eq:r_dot}-\ref{eq:s_dot}) for a network of IK neurons with neuron-specific recovery variables $u_i$, as defined by Eq.\eqref{eq:u_i}.
Following the approach of \cite{chen_exact_2022}, we showed that the mean field dynamics of the average recovery variable $u$ as defined by Eq.\eqref{eq:u} are coupled to the average membrane potential $v$ and average firing rate $r$ of the population (see Eq.\eqref{eq:u_dot}).
This result is equivalent to the result obtained in \cite{guerreiro_exact_2022}, where the mean-field equations were derived using an adiabatic approximation based on the assumption that the dynamics of the recovery variables $u_i$ are slow in comparison to the dynamics of the membrane potentials $v_i$.

Strikingly, the mean-field equation for $u$ is identical to the mean-field equation derived in \cite{gast_effects_2022} for spiking neural networks in which all neurons share a single global recovery variable $u$.
The dynamic equations of the spiking neural network considered in \cite{gast_effects_2022} are given as

\begin{align}
    C \dot v_i &= k (v_i - v_r)(v_i - v_{\theta,i}) - u + I + g s (E-v_i), \label{eq:v_global}\\
    \tau_u \dot u &= - u + \frac{b}{N} \sum_{j=1}^N (v_j - v_r) + \frac{\tau_u \kappa}{N} \sum_{j=1}^N \delta(v_j-v_p), \label{eq:u_global}
\end{align}

where $s$ is still given by Eq.\eqref{eq:s}.
Although both the network with neuron-specific recovery variables $u_i$ and the network with a global recovery variable $u$ produce the same mean-field equations, it is likely that their dynamics are not identical.
To examine how spiking neural networks with neuron-specific vs. global recovery variables differ in their dynamics, and to determine how both spiking models differ from their mean-field approximation, we compare the dynamics of both the spiking neural network given by eqs.(\ref{eq:v_i}-\ref{eq:u_i}) and the spiking neural network given by eqs.(\ref{eq:v_global}-\ref{eq:u_global}) to the dynamics predicted by the mean-field model of eqs. (\ref{eq:r_dot}-\ref{eq:s_dot}).

\section{Mean-field modeling of adaptation-induced bursting}

\begin{figure*}
    \centering
    \includegraphics[width=1.0\textwidth]{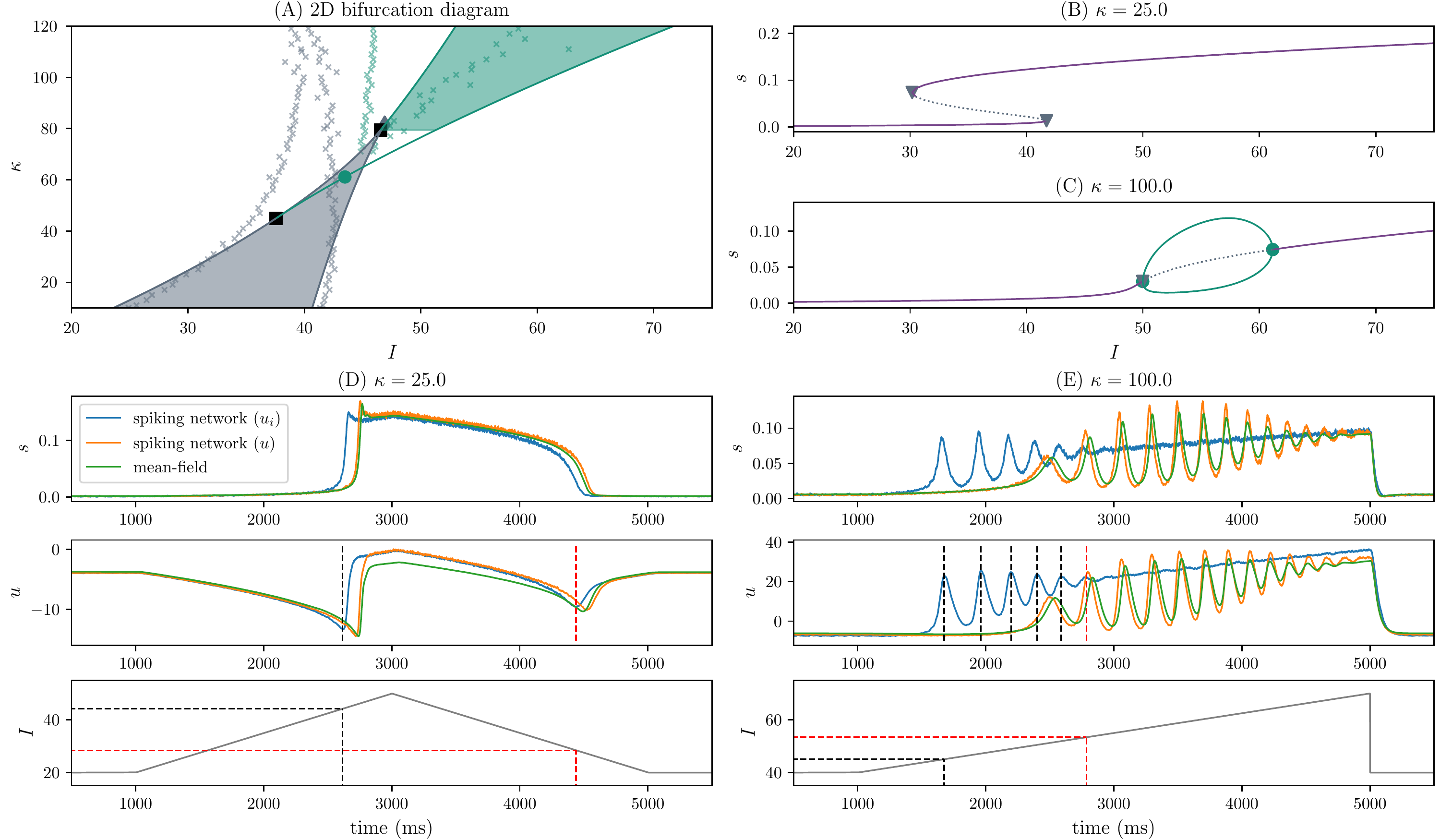}
    \caption{\textit{Strong spike-frequency adaptation reduces mean-field accuracy.} \textbf{(A)} Bifurcations in the 2D parameter space spanned by the spike-frequency adaptation strength $\kappa$ and the input current $I$. Green (grey) regions depict synchronized-oscillatory (asynchronous, bistable) regimes. Green (grey) solid lines depict curves of Hopf bifurcation (fold bifurcation) solutions. Black squares, grey diamonds, and green circles represent Bogdanov-Takens, cusp, and generalized Hopf bifurcations, respectively. Grey (green) $x$ markers depict fold (Hopf) bifurcations identified in the dynamics of a spiking neural network with neuron-specific recovery variables $u_i$. \textbf{(B-C)} Solution for the mean-field variable $s$ as a function of $I$ for two different spike frequency adaptation strengths. Purple solid lines represent stable steady-state solutions; grey dotted lines represent unstable steady-state solutions. Green solid lines represent the minima and maxima of limit cycle solutions. \textbf{(D-E)} Dynamics of the mean-field model (green), the spiking neural network with neuron-specific recovery variables $u_i$ (blue), and the spiking neural network with a global recovery variable $u$ (red), for two different values of the spike frequency adaptation strength $\kappa$. The first and second row depict the synaptic activation dynamics and recovery variable dynamics, respectively (both averaged across neurons), whereas the last row depicts the applied input current $I$ as a function of time. Note that for illustration purposes, input currents shown here increase over a much faster timescale than used to identify approximate bifurcation points in (A), and that the input current in (D) was decreased (increased) in the second (first) half of the trial to locate the fold bifurcations that lie on the left (right) solution branch of the bistable region in (A). Vertical dashed lines represent the troughs in $u$ used to locate fold and Hopf bifurcation points for the input parameter $I$. Horizontal dashed lines indicate the values of $I$ at which fold or Hopf bifurcations were located.}
    \label{fig:sfa}
\end{figure*}

Our first assumption in deriving the mean-field model (\ref{eq:r_dot}-\ref{eq:s_dot}) is that spike-frequency adaptation is small, i.e. $u \gg \kappa$ for any $v_{\theta}$.
Here, we examine how well the predictions of the mean-field theory capture the dynamics of the spiking neural network when $\kappa$ is systematically varied.
To this end, we performed a bifurcation analysis of the mean-field model over input current $I$ and adaptation parameter $\kappa$, using PyRates \cite{gast_pyratespython_2019} and Auto-07p \cite{doedel_auto-07p:_2007}, and compared it to numerical approximations of the bifurcation structure of the spiking neural network.

To locate the bifurcation points for the spiking neural network, we performed numerical integration of the network eqs.(\ref{eq:v_i}-\ref{eq:s}) for an all-to-all coupled population of $N = 10000$ neurons over a time interval of $20 \text{ s}$, much greater than the longest time constant of the model neurons, $C/k ~ 142\textrm{ ms}$).
We  used the explicit Euler method with an integration step size of $0.001 \text{ ms}$.
Over the course of the integration interval, we slowly ramped up the background current $I(t)$ from $20 \text{ pA}$ to $70 \text{ pA}$ in the first $10 \text{ s}$ and then linearly decreased it back to $20 \text{ pA}$ in the second $10 \text{ s}$, resulting in a rate of change of $5.0 \text{ pA/s}$.
We used the troughs of the recovery variable $u_i$ averaged over the population, $\langle u_i \rangle_i$, to locate fold and Hopf bifurcations as a function of the input current $I$.

Fig.\ref{fig:sfa}D and E depict representative dynamics of the average recovery variable and the background current $I$ as used for locating fold (in D) and Hopf (in E) bifurcations; a shorter time interval was used in these figures for readability.
For simplicity, we use $u$ to refer to the population average recovery variables of each of the three models that we compare in Fig.\ref{fig:sfa}D and E, i.e. the global recovery variable of the mean field model given by Eq.\eqref{eq:u_dot}, the global recovery variable given by Eq.\eqref{eq:u_global}, and the average of the neuron-specific recovery variables $\langle u_i \rangle_i$ with $u_i$ given by Eq.\eqref{eq:u_i}.
Note that while fold bifurcations are identified as broad, single troughs in $\langle u_i \rangle_i$, Hopf bifurcations are located at the borders of intervals with multiple, more narrow troughs.
We repeated this procedure for multiple values of the spike-frequency adaptation strength $\kappa$ to approximate the fold and Hopf bifurcation curves in the 2D parameter plane spanned by $I$ and $\kappa$.
All other model parameters were set to the values reported in Tab.\ref{tab:rs}.

\begin{table}[h]
\caption{Model parameters for a regular-spiking IK neuron\label{tab:rs}}
\begin{ruledtabular}
\begin{tabular}{cccc}
\textrm{Parameter} & \textrm{Value} & \textrm{Parameter} & \textrm{Value}\\
\colrule
$C$ & $100$ & $k$ & $0.7$\\
$v_r$ & $-60$ & $\bar v_{\theta}$ & $-40$\\
$g$ & $1$ & $E$ & $0$\\
$\tau_u$ & $33.33$ & $\tau_s$ & $6.0$\\
$\kappa$ & $10$ & $b$ & $-2.0$\\
$J$ & $15$ & $N$ & $10000$\\
$v_p$ & $1000$ & $v_0$ & $-1000$\\
\end{tabular}
\end{ruledtabular}
\end{table}

Fig.\ref{fig:sfa}A shows that $\kappa$ controls whether the spiking neural network expresses a bistable or an oscillatory regime: the former exists for small values of $\kappa$, whereas the latter requires relatively large values of $\kappa$.
As expected, we find that the accuracy of the mean-field model is reduced when $\kappa$ is increased and the $u \gg \kappa$ assumption of \cite{chen_exact_2022} is violated.
The location of the fold bifurcations predicted by the mean-field theory matches the location of the fold bifurcations estimated from the spiking neural network dynamics for $1 < \kappa < 40$.
However, the larger $\kappa$ becomes, the stronger is the deviation between the bifurcation curves calculated from the mean-field model and the ones extracted from the dynamics of a spiking neural network with neuron-specific recovery variable $u_i$ (see Fig.\ref{fig:sfa}A).
Note that the average values of $u$ do not meet the condition $u \gg \kappa$ for most values of $\kappa$ depicted in Fig.\ref{fig:sfa}A, as can be seen in the middle row of Fig.\ref{fig:sfa}D and E.

Nevertheless, we find that even for larger values of $\kappa$ the spiking neural network exhibits a bifurcation structure that is qualitatively similar to that of the mean-field model.
The bistable regime is most pronounced at small values of $I$ and $\kappa$, and the oscillatory regime emerges for higher values of $I$ and $\kappa$.
Violations of the small spike-frequency adaptation assumption merely lead to a shift of the bifurcation curves in parameter space; this shift increases as $\kappa$ increases.
Finally, Fig.\ref{fig:sfa}D and E demonstrate that the mean-field predictions are in better agreement with the dynamics of the spiking neural network governed by a global recovery variable $u$, as described by eqs.(\ref{eq:v_global}-\ref{eq:u_global}).
This reflects the fact that the mean-field model effectively assumes that the fluctuations of the  $u_i$ variable across neurons are negligible.
While this assumption naturally holds in the spiking neural network with a global recovery variable $u$, it does not necessarily hold for networks with individual recovery variables $u_i$.
This is particularly the case in networks with spike threshold or input current heterogeneity.
Neurons with different spike thresholds will differ in their individual firing rates, which causes heterogeneity in the recovery variables $u_i$ via the dependence of $u_i$ on those firing rates, scaled by $\kappa$ (see Eq.\eqref{eq:u_i}).
We conclude that spiking neural networks with neuron-specific recovery variables $u_i$ behave qualitatively similarly to spiking neural networks with a global recovery variable $u$, but that the quantitative agreement between the two becomes worse as $\kappa$ increases. 

\section{Mean-field modeling of different spike waveforms}

Another factor limiting the applicability of the mean-field model is the assumption that $v_p \rightarrow \infty$ and $v_0 \rightarrow -\infty$, namely that a spike is emitted as the membrane potential approaches a peak of $\infty$, and that following a spike the membrane potential resets to $-\infty$.
These assumptions were necessary for the analytic derivation of the mean-field equations.
However, the variety of firing patterns that the IK neuron model is able to exhibit depends on finite values of $v_p$ and $v_0$ \cite{izhikevich_simple_2003,izhikevich_dynamical_2007}.

In this section, we examine the mismatch between the mean-field model and spiking neural network dynamics given finite, realistic peak and reset potentials.
To correct for this mismatch, we introduce an input rescaling factor $I^*$ that allows the mean-field model to be adapted to better match the observed dynamics of spiking neural networks with finite spike resetting parameters.

\subsection{Deriving the relationship between peak/reset potential values and firing rate of the IK model neuron}

\begin{figure*}
    \centering
    \includegraphics[width=1.0\textwidth]{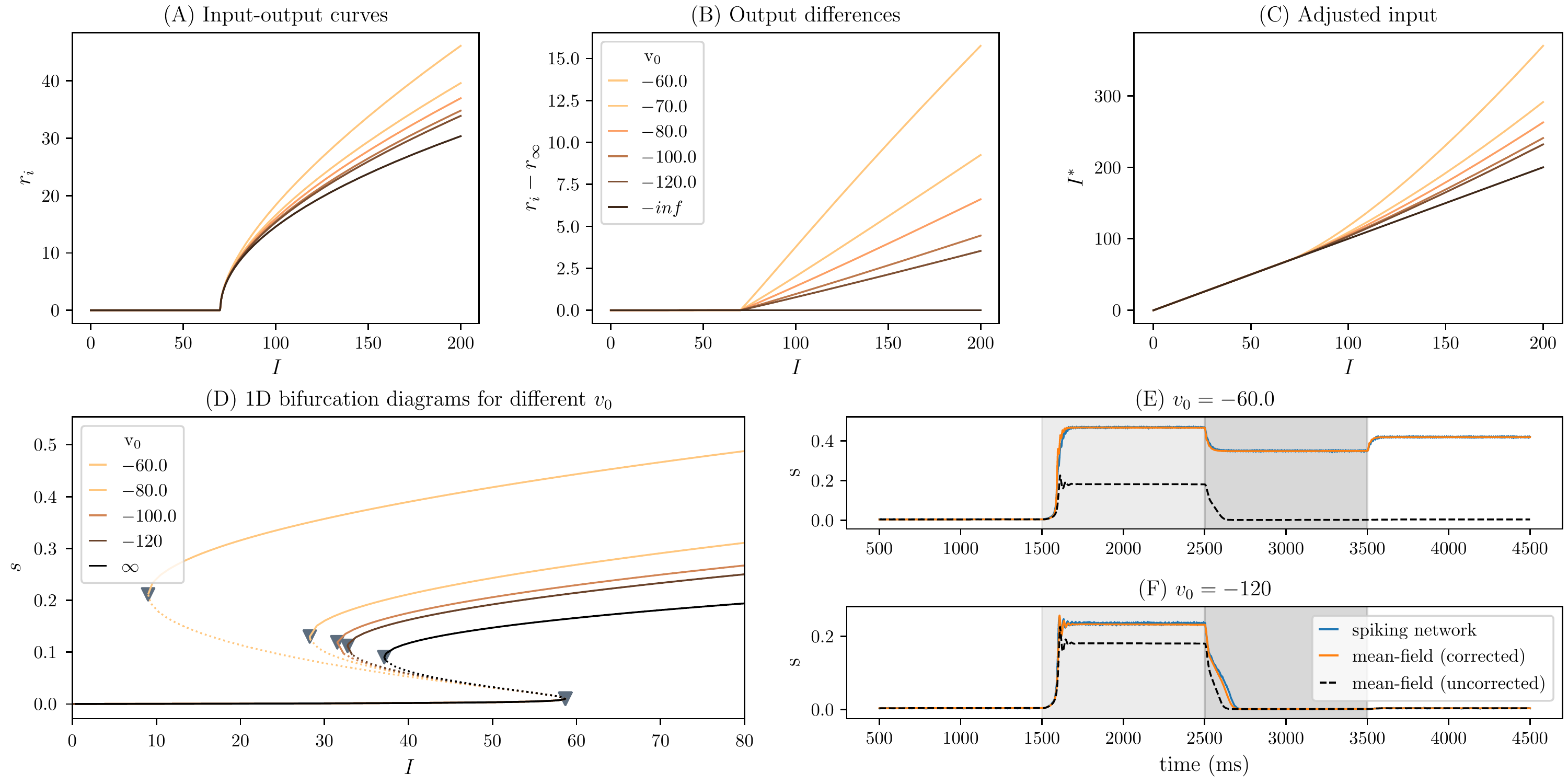}
    \caption{\textit{Firing rates under different spike reset conditions.} \textbf{(A)} Steady-state firing rates $r_i$ of single neurons as a function of the input $I$ for different reset potentials $v_0$. Color code as in B. \textbf{(B)} Differences between the steady-state firing rates of a neuron with $v_0 \rightarrow -\infty$ as assumed in the mean field model and neurons with different finite values of the reset potential $v_0$. \textbf{(C)} Adjusted input $I^*$ as a function of $I$ for different reset potentials as per Eq.\eqref{eq:I_corr}. \textbf{(D)} Steady-state solution for the  mean-field variable $s$ as a function of the input $I$. Grey triangles represent fold bifurcations. Solid (dotted) lines represent stable (unstable) solutions. \textbf{(E-F)} Synaptic activation dynamics $s$ for the spiking neural network, and for both uncorrected and corrected mean-field models, for two values of $v_0$.}
    \label{fig:spiking}
\end{figure*}

We first analyze the impact of $v_p$ and $v_0$ on the dynamics of a single IK neuron.
Neither parameter enters into Eq.\eqref{eq:v_i}; $v_p$ and $v_0$ only affect the IK neuron dynamics in the spiking regime, where spike-triggered resetting of the membrane potential takes place.
It is in this regime that we examine the effect of $v_p$ and $v_0$ on the dynamics of a single IK neuron.
The adiabatic approximation that $u_i$ changes infinitesimally slowly with respect to $v_i$ leads to an analytical solution to Eq.\eqref{eq:v_i}

\begin{equation}
    v_i(t) = \frac{1}{2}\Big[\sqrt{\mu_i} \tan\Big(\frac{kt\sqrt{\mu_i}}{2C} + \tan^{-1}\Big(\frac{2v_i(t_0) - \frac{\alpha}{k}}{\sqrt{\mu_i}}\Big)\Big) + \frac{\alpha}{k}\Big], \label{eq:vi_sol}
\end{equation}
where
\begin{equation}
    \mu_i = \frac{4 \beta}{k} - \Big(\frac{\alpha}{k}\Big)^2 \label{eq:mu_i}
\end{equation}

can be interpreted as a lumped sum of input currents to the neuron, with  $\alpha$ and $\beta$ given by Eq.\eqref{eq:alpha} and Eq.\eqref{eq:beta}, respectively.
We assume $\mu_i > 0$, which is equivalent to assuming that the neuron is in a spiking regime.
For more detailed descriptions of the adiabatic approximation and how it can be used to absorb a spike-frequency adaptation variable into the membrane potential dynamics of a spiking neuron, see \cite{gigante_diverse_2007,gast_mean-field_2020,guerreiro_exact_2022}.

Based on Eq.\eqref{eq:vi_sol}, the spiking frequency of an IK neuron receiving a positive lumped input current $\mu > 0$ can be calculated by setting $v_i(t_0) = v_0$ and solving for the time $t$ it takes for $v_i(t)$ to reach $v_p$, yielding

\begin{equation}
    r_i = \frac{k\sqrt{\mu_i}}{2C\gamma_i}, \label{eq:r_i}
\end{equation}
where
\begin{equation}
    \gamma_i = \tan^{-1} \Big(\frac{2v_p-\frac{\alpha}{k}}{\sqrt{\mu_i}}\Big) - \tan^{-1}\Big(\frac{2v_0-\frac{\alpha}{k}}{\sqrt{\mu_i}}\Big). \label{eq:gamma_i}
\end{equation}

Eqs.(\ref{eq:r_i}-\ref{eq:gamma_i}) establish a functional relationship between the firing rate $r_i$ of a single IK neuron and the spike reset condition defined via $v_p$ and $v_0$. 

\subsection{Mean-field correction for spike resetting}

To obtain a revised set of mean-field eqs.(\ref{eq:r_dot}-\ref{eq:s_dot}) that correct for the effects of finite peak and reset potentials, we used the limits $v_p \rightarrow \infty$ and $v_0 \rightarrow -\infty$. 
In this limit, $\gamma_i \rightarrow \pi$ and Eq.\eqref{eq:r_i} simplifies to

\begin{equation}
    r_{\infty} = \frac{k \sqrt{\mu_i^*}}{2\pi C}, \label{eq:r_infty}
\end{equation}
where $\mu_i^*$ is defined as
\begin{equation}
    \mu_i^* = \frac{4 \beta^*}{k} - \Big(\frac{\alpha}{k}\Big)^2. \label{eq:mu_star}
\end{equation}
In Eq.\eqref{eq:mu_star}, $\beta^* = k v_r v_{\theta} + g s E - u + I^*$, and $I^*$ is an "adjusted" extrinsic input current that can be different from $I$.
The differences between $r_i$ and $r_{\infty}$ in the absence of the adjustment factor are depicted for different values of $v_0$ in Fig.\ref{fig:spiking}A and B for $I^* = I$

These differences in the output firing rates of single neurons will cause a corresponding mismatch between the firing rates predicted by the mean-field theory and those of a spiking neural network with $v_p \ll \infty$ and/or $v_0 \gg -\infty$.
At the single cell level, the difference in firing rates between $r_i$ and $r_{\infty}$ for $v_p < \infty$ and $v_0 > -\infty$ can be corrected by choosing the adjusted extrinsic input as

\begin{equation}
     I^* = 
     \begin{cases}
     \frac{\pi^2 k \mu}{4\gamma^2} + \frac{\alpha^2}{4k} + u_i - k v_r v_{\theta} - g s E & \text{if } \mu > 0,\\
     I, & \text{otherwise.}
     \end{cases}
     \label{eq:I_corr}
\end{equation}

Fig.\ref{fig:spiking}C shows the resulting relationship between $I^*$ and $I$.
It reveals that $I^* \geq I$ is required to achieve $r_{\infty} = r_i$ when $I$ is large enough to elicit spiking, and that the magnitude of the difference grows with $I$ and with $v_0$, which shapes $I^*$ through its contribution to $\gamma$.
The piecewise structure of Eq.\eqref{eq:I_corr} preserves a monotonic and continuous relationship between $I$ and $I^*$.
Continuity follows from evaluating $\lim_{\mu \to 0^+} I^*$.
In this limit, the term $\frac{\pi^2 k \mu}{4\gamma^2}$ in Eq.\eqref{eq:I_corr} $\to 0$ and $I^* \to I$.

To incorporate this input adjustment into the mean-field theory, we derive the mean-field equations for a network of globally coupled IK neurons where the membrane potential of the $i^\textrm{th}$ neuron evolves according to

\begin{equation}
     C \dot v_i = k (v_i - v_r)(v_i - v_{\theta}) - u_i + I^* + g s (E-v_i), \label{eq:v_corr}\\
\end{equation}

instead of Eq.\eqref{eq:v_i}.
We use a first-order approximation to $\frac{\partial}{\partial v_{\theta}} I^*$, which allows us to simplify Eq.\eqref{eq:I_corr} by replacing $v_{\theta}$ by $\bar v_{\theta}$ and assuming that both $\mu$ and $\gamma$ are functions of $\bar v_{\theta}$ instead of $v_{\theta}$.
This approximation amounts to setting the corrected input $I^*$ to all neurons to that of the average neuron of the network, disregarding any potential effects in the mean-field dynamics due to fluctuations in $I^*$ caused by fluctuations in $ v_{\theta}$.
Under this assumption, the mean-field equations can be derived as outlined in Section \ref{sec:derivation}, to obtain

\begin{align}
    C \dot{r} = &\frac{\Delta_v k^2}{\pi C} (v-v_r) + r(k(2 v - v_r - \bar v_{\theta}) - g s), \label{eq:r_dot_c}\\
    C \dot{v} = &k v(v - v_r - \bar v_{\theta}) - \pi C r(\Delta_v + \frac{\pi C}{k} r) \label{eq:v_dot_c}\\
    &+ v_r \bar v_{\theta} - u + I^* + g s (E-v), \nonumber \\
    \tau_u \dot{u} = &b(v-v_r) - u + \tau_u \kappa r, \label{eq:u_dot_c}\\
    \tau_s \dot s = &-s + \tau_s J r, \label{eq:s_dot_c}
\end{align}

with $I^*$ given by Eq.\eqref{eq:I_corr} with $v_{\theta} \to \bar v_{\theta}$.
Importantly, the continuous nature of $I^*$ allows for the application of methods from dynamical systems theory such as numerical parameter continuation.
As shown in Fig.\ref{fig:spiking}D-F, the correction term leads to a substantially improved agreement between the mean-field theory and the spiking neural network dynamics.
Note, however, that the striking agreement between mean-field theory and spiking neural network dynamics shown in Fig.\ref{fig:spiking}E and F holds for the optimal condition of a single population of IK neurons with $b = 0$ and $\kappa = 0$ (all other parameters were chosen according to Tab.\ref{tab:rs}).
Under these conditions, any potential mismatch that might arise due to the adiabatic approximation we used to obtain Eq.\eqref{eq:vi_sol} can be neglected.

It should also be noted that the correction term $I^*$ becomes less accurate when $v_0 \geq v_r$,  that is when the reset potential after spiking is above the resting membrane potential, as is the case in some bursty spiking neurons.
Under this condition, spike resetting affects not only the firing rate but also sub-threshold dynamics.
Since $I^*$ as given by Eq.\eqref{eq:I_corr} only applies a correction when $\mu > 0$, no correction is applied in these sub-threshold regimes.

\begin{figure*}
    \centering
    \includegraphics[width=1.0\textwidth]{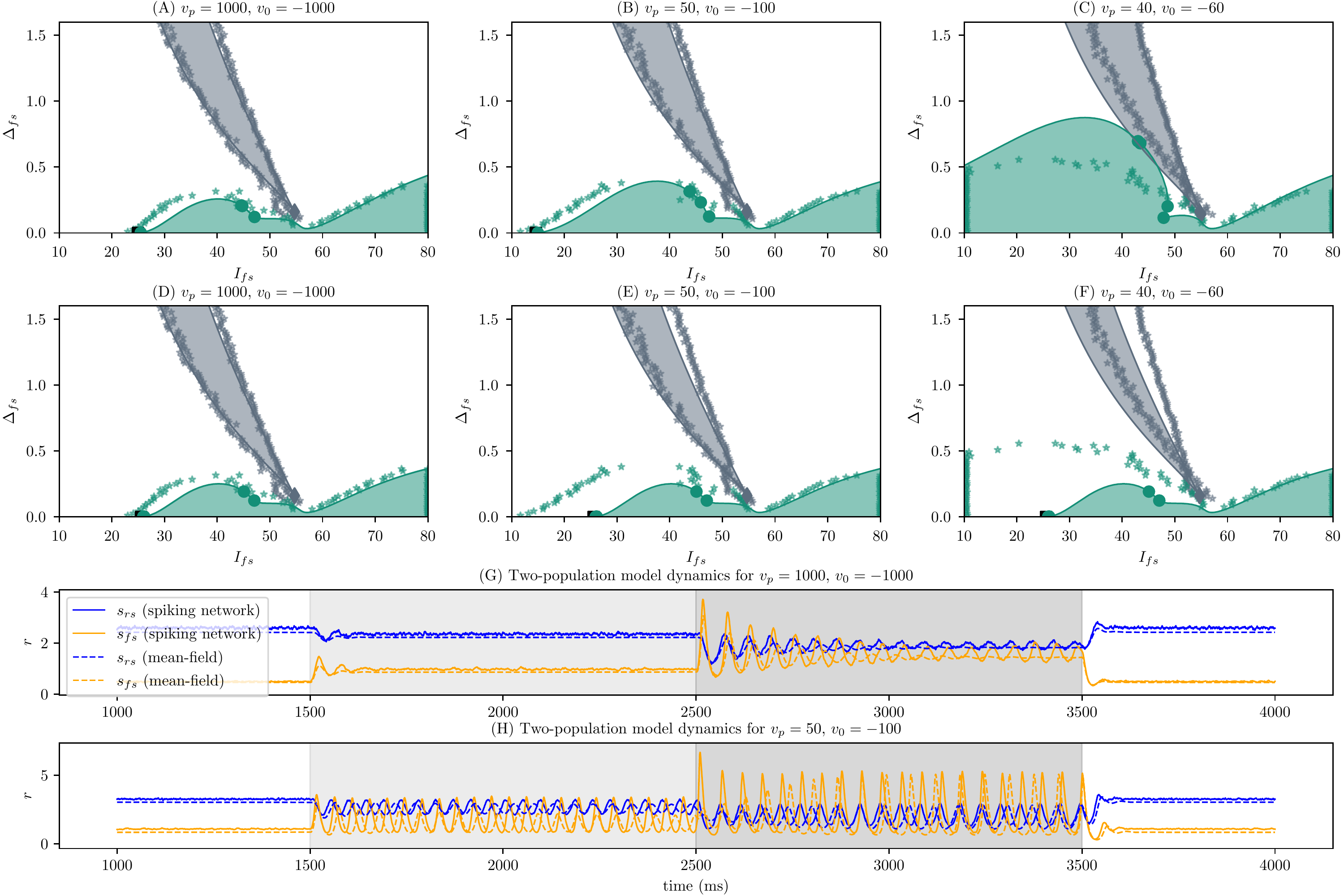}
    \caption{\textit{Bifurcation structure and network dynamics in an excitatory-inhibitory network for different spike reset conditions.} \textbf{A-C} 2D bifurcation diagrams in the plane of fast-spiking spike threshold heterogeneity $\Delta_{fs}$ and fast-spiking neuron input $I_{fs}$ for three different spike reset conditions. Regions of parameters space depicted in green (grey) represent synchronized-oscillatory (asynchronous-bistable) regimes as predicted by the corrected mean-field model. Solid green (grey) lines depict the Hopf (fold) curves predicted by the corrected mean-field model. Grey diamonds represent cusp bifurcations and green circles represent generalized Hopf bifurcations. Green (grey) $x$ markers depict the locations of Hopf (fold) bifurcations from the spiking neural network dynamics. \textbf{(D-F)} Same as (A-C), except that mean-field predictions follow from the uncorrected mean-field model. \textbf{(G-H)} Firing rate dynamics of the regular {\it rs} (blue) and fast {\it fs} (orange) spiking neurons for two different spike reset conditions. Solid (dashed) lines represent the mean-field model dynamics. The input to the fast-spiking neurons was stepped from $I_{fs} = 20$ (no shading) to $I_{fs} = 30$ (light grey shading), to $I_{fs} = 40$ (dark grey shading).}
    \label{fig:eic}
\end{figure*}

\subsection{Effects of spike resetting on the dynamics of a two-population model}

To test whether the corrected $I^*$ can also improve the agreement between mean-field theory and spiking neural network dynamics for finite values of $v_p$ and $v_0$ under less optimal conditions, we considered a network of interacting regular-spiking and fast-spiking neurons, using the model equations and parameters reported in \cite{gast_effects_2022}.
We compared the mean-field predictions of the uncorrected and corrected mean-field models with the spiking neural network dynamics of this two-population network for three different spike reset conditions: $v_p = 1000$ and $v_0 = -1000$, $v_p = 50$ and $v_0 = -100$, and $v_p = 40$ and $v_0 = -60$.
Again, we used numerical bifurcation analysis to identify the bifurcation structure of the mean-field model in the 2D parameter space spanned by the background current to the fast-spiking neuron population $I_{fs}$ and the width of the spike threshold distribution across fast-spiking neurons $\Delta_{fs}$.
To identify the location of the fold and Hopf bifurcations in the two-population spiking neural neuron network, we used the method described in the previous section.

The comparison between Fig.\ref{fig:eic}A-C and Fig.\ref{fig:eic}D-F reveals that the corrected mean-field model predicts synchronized oscillations in the dynamics of the spiking neural network more accurately than the uncorrected mean-field model.
This agreement is shown in Fig.\ref{fig:eic}G and Fig.\ref{fig:eic}H, for spiking neural networks with $v_p = 1000$ and $v_0 = -1000$ or $v_p = 50$ and $v_0 = -100$, respectively.
Furthermore, we find that the bifurcation structure of the spiking neural network with realistic spike resetting, $v_p = 50$ and $v_0 = -100$, follows the prediction of the corrected mean-field model (see Fig.\ref{fig:eic}B), while the IK network with less realistic spike resetting, $v_p = 1000$ and $v_0 = -1000$, shows a bifurcation structure closer to that of the uncorrected mean-field model (see Fig.\ref{fig:eic}D). 
As expected, the correction becomes less necessary as the absolute values of $v_p$ and $v_0$ become unrealistically large. 

Figs.\ref{fig:eic}B, C, and H further reveal that the correction term becomes less accurate as the absolute values of $v_p$ and $v_0$ decrease.
The bifurcation structure predicted by the corrected mean-field model accurately captures the dynamics of the corresponding spiking neural network for $v_p = 50$ and $v_0 = -100$, but the mean-field predictions for $v_p = 40$ and $v_0 = -60$ overestimate the areas of the oscillatory and bistable regions in the 2D parameter space spanned by $\Delta_{fs}$ and $I_{fs}$.
This discrepancy arises because we have assumed that the firing thresholds $v_{\theta}$ of the neurons in the spiking neural network followed a heavy-tailed distribution.
In the regime where the absolute values of $v_p$ and $v_0$ are small, the effective range of values for $v_{\theta}$ becomes narrower, as it cannot be the case that $v_{\theta} > v_p$ or $v_{\theta} < v_0$. 
In the following section, we investigate the general issue of defining a probability distribution $p(v_{\theta})$ on a restricted domain, and its implications for the mean-field model.

We conclude that inaccuracies between spiking network and mean-field theory arising from finite spike resetting conditions can be accounted for by introducing a corrected input term in the mean-field model.
This correction term provides a substantially improved fit of the spiking network dynamics for $v_0 < v_r$ and $v_p$ large enough that the cumulative probability density $\int_{v_{\theta} = v_p}^{v_{\theta} = \infty} p(v_{\theta}) d v_{\theta}$ is sufficiently small.

\section{Mean-field effects of truncated distributions}

\begin{figure}
    \centering
    \includegraphics[width=1.0\columnwidth]{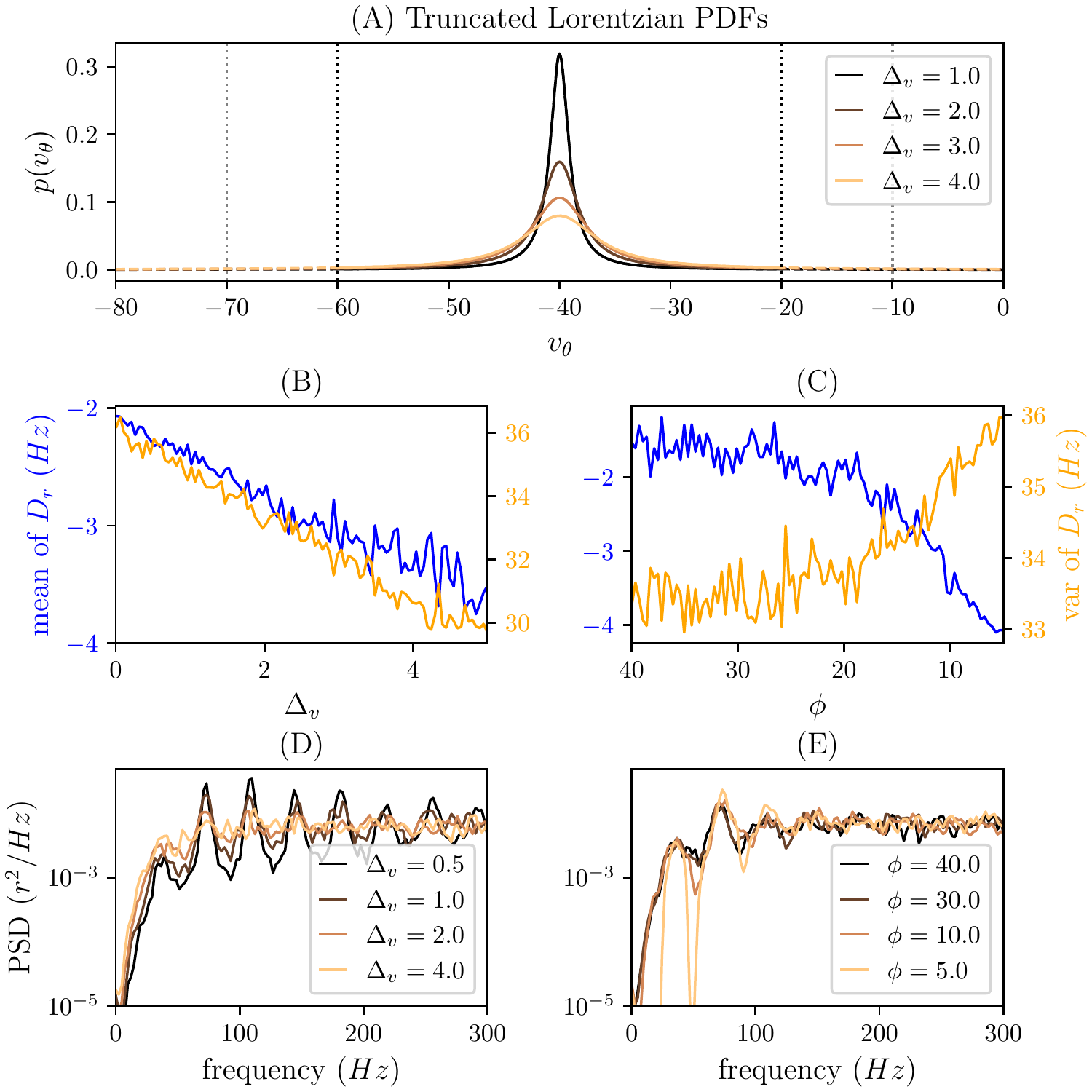}
    \caption{\textit{Effects of a truncated Lorentzian distribution of spike thresholds on the mean-field dynamics.} \textbf{(A)} Probability density as a function of the spiking threshold $v_{\theta}$ for Lorentzian distributions with $\bar v_{\theta} = -40$ and different widths $\Delta_v$. The transition from solid to dashed lines illustrates the truncation. Dotted vertical lines represent two different values of the truncation threshold $\phi$. \textbf{(B-C)} Mean and variance of the difference $D_r$ between the average firing rates of the mean-field model vs. the spiking network as a function of the Lorentzian width $\Delta_v$ (B) and the cutoff threshold $\phi$ (C). \textbf{(D-E)} Power-spectral density (PSD) of the average firing rate dynamics in the spiking network for different spiking thresholds and truncation thresholds, respectively.}
    \label{fig:lorentz}
\end{figure}

Here we consider the effects of assuming that the spiking thresholds $v_{\theta}$ in the spiking neural network follow a Lorentzian distribution across neurons, with probability density function given by Eq.\eqref{eq:p_lorentzian}.
This assumption allows for a particularly strong reduction in the dimensionality of the mean-field equations, but it introduces a heavy-tailed distribution defined over the entire domain $[-\infty, \infty]$.
The spiking threshold $v_{\theta}$ is a membrane potential, a continuous variable that can, in principle, span the unbounded domain $[-\infty, \infty]$, but the symmetry of the term $(v_i - v_r)(v_i - v_{\theta})$ in Eq.\eqref{eq:v_i}, indicates that $v_{\theta}$ effectively becomes $v_r$ and vice versa if $v_r > v_{\theta}$. 
So, $v_{\theta}$ should be bounded from below: $v_{\theta} > v_r > v_0$. 
It is evident from Eq.\eqref{eq:v_i} that $v_{\theta}$ is also bounded from above, as $v_i$ could never diverge to produce a spike if $v_{\theta} > v_p$.

Therefore in practice $v_r < v_{\theta} < v_p$, and a truncated Lorentzian distribution must be used at the level of the spiking neural network, such that $v_r < v_{\theta,i} < v_p \ \text{ } \forall i \in [1,2,...,N]$.
We use a truncated Lorentzian that retains the symmetry of the original Lorentzian

\begin{equation}
    p^*(v_{\theta}) = 
    \begin{cases}
        g_{\phi} p(v_{\theta}) & \text{if } \bar v_{\theta} - \phi < v_{\theta} < \bar v_{\theta} + \phi,\\
        0 & \text{otherwise,}
    \end{cases}
    \label{eq:lorentzian_truncated}
\end{equation}

where $\phi$ is the truncation threshold, $g_{\phi}$ is a normalization constant that enforces $\int_{-\infty}^{\infty} p^*(v_{\theta}) d v_{\theta} = 1$, and $p(v_{\theta})$ is given by Eq.\eqref{eq:p_lorentzian}.
We used such a truncated Lorentzian with $\phi = v_r$ for all spiking neural network results reported above.
Since the mean-field derivation requires the use of the full Lorentzian distribution, 
the mean-field model based on a full Lorentzian distribution over the spiking thresholds can only approximate the spiking model based on a truncated Lorentzian distribution over this model parameter.

We now examine how this approximation affects the agreement between mean-field theory and spiking neural network dynamics.
To this end, we compared the average firing rate dynamics of the mean-field model and the spiking network of a population of coupled regular-spiking neurons for different truncation conditions.
We systematically varied either the width of the spike threshold distribution $\Delta_v$ or the truncation threshold $\phi$, and calculated the firing rate difference $D_r = r - \langle r_i \rangle_i$. 
Mean-field and spiking neural network dynamics were obtained from the numerical integration of eqs.(\ref{eq:r_dot_c}-\ref{eq:s_dot_c}) and eqs.(\ref{eq:v_i}-\ref{eq:s}), respectively, using the parameters from Tab.\ref{tab:params2}.

\begin{table}[h]
\caption{Model parameters for IK neurons used to study the effects of truncated Lorentzian spike threshold distributions \label{tab:params2}}
\begin{ruledtabular}
\begin{tabular}{cccc}
\textrm{Parameter} & \textrm{Value} & \textrm{Parameter} & \textrm{Value}\\
\colrule
$C$ & $100$ & $k$ & $0.7$\\
$v_r$ & $-80$ & $I$ & $250$\\
$g$ & $1$ & $E$ & $0$\\
$\tau_u$ & $33.33$ & $\tau_s$ & $6.0$\\
$\kappa$ & $20$ & $b$ & $-2.0$\\
$J$ & $15$ & $N$ & $10000$\\
$v_p$ & $1000$ & $v_0$ & $-1000$\\
\end{tabular}
\end{ruledtabular}
\end{table}

The results of these calculations are shown in Fig.\ref{fig:lorentz}.
As expected, either increasing the width of the spike threshold distribution $\Delta_v$ or reducing the truncation threshold $\phi$ leads to an increased average firing rate difference ($D_r$) between mean-field prediction and spiking network dynamics (Fig.\ref{fig:lorentz}B-C).  
In both cases the cumulative probability density in the truncated tails is increased, rendering the mean-field assumption less accurate.
Interestingly, increases in $\Delta_v$ led to a decreased variance of $D_r$, whereas decreases in $\phi$ led to an increased variance of $D_r$.
As shown in Fig.\ref{fig:lorentz}D and E, this is due to the desynchronization of a neural population with inhibitory coupling that is caused by increased spike threshold heterogeneity \cite{gast_effects_2022}.  
Both decreasing the spike threshold distribution width and decreasing the truncation threshold result in a more homogeneous spiking network and thus cause increased synchrony in its firing rate fluctuations.
As can be seen from the difference between Figs.\ref{fig:lorentz}D and E, the spectral properties of these synchronous network dynamics depend on the particular shape of the spike threshold distribution. 

For a more detailed discussion of the relationship between population dynamics and spike threshold heterogeneity, see \cite{gast_effects_2022}.
Importantly, we find that the magnitude of the average firing rate differences between mean-field model and spiking network is small compared to the firing rate fluctuations in the spiking network; this discrepancy is thus likely to have little effect on qualitative aspects of network dynamics.
This is as seen in Fig.\ref{fig:eic}A-C, where increases in the width of the spike threshold distribution of fast-spiking neurons $\Delta_{fs}$ (y axis) do not affect the accuracy of the mean-field predictions.   

\section{Conclusion}

The spiking activity of neurons is shaped by their underlying electrophysiological properties; different cell types typically exhibit dramatically different spiking responses to the same input.
To understand the computational consequences of this diversity, we must study its effect at the level of neural population dynamics.
Approaches such as mean-field modeling provide insight into the emergent dynamics of neural populations, but these models most commonly treat all neurons as identical copies of each other and omit physiological properties differentially associated with known cell types in the brain.
Here we have presented a mean-field model of a network of coupled Izhikevich (IK) neurons with biophysiological state variables and parameters, an approach that allows us to predict how neural population dynamics are shaped by the distinct response properties of individual neurons.

A key advantage of IK model neurons is that the parameters and state variables of the model neurons, such as the membrane capacitance, the membrane potential, or the maximum conductance of synapses, are based on electrophysiological properties that can be measured directly \cite{izhikevich_which_2004,izhikevich_dynamical_2007}.
Through the tuning of these parameters, the IK model can represent various neuron and synapse types, and thus account for different sources of neural heterogeneity in the brain \cite{izhikevich_simple_2003,guerreiro_exact_2022}. 
These features render our mean-field model particularly suited for interpreting neural recordings and developing large-scale models of multiple interacting neuron types \cite{izhikevich_large-scale_2008}.
Our work contributes to such efforts by providing a mean-field model that links single cell properties to population-level dynamics, thus helping to bridge different scales of brain organization \cite{deco_dynamic_2008,coombes_large-scale_2010,engel_intrinsic_2013,vohryzek_understanding_2022}. 

Our model also introduces a novel approach to account for heterogeneity in neuron spike thresholds, a property of neuron populations that has been well characterized experimentally (e.g., \cite{wang_anatomical_2004,yang_distinct_2013,neske_contributions_2015}.)
As we demonstrate in \cite{gast_effects_2022}, the degree of variance in spiking thresholds across a neural population has strong effects on the dynamic regimes that these populations exhibit.
In \cite{gast_effects_2022}, we found these effects to determine the dynamic regimes of excitatory-inhibitory circuits that represent key elements of cortical organization \cite{potjans_cell-type_2014,schwalger_towards_2017,jonke_feedback_2017}.
In this work, we provided a detailed analysis of the simplifying assumptions required to derive the mean-field model for populations of heterogeneous spiking neurons and the biases that these simplifying assumptions introduce in the predicted mean-field dynamics.

For the derivation of the mean-field model, we built upon previous work that derived mean-field equations for spiking neural networks \cite{luke_complete_2013,montbrio_macroscopic_2015} in a manner that avoids often invoked asynchronous firing \cite{amit_model_1997,vreeswijk_chaotic_1998,brunel_dynamics_2000,el_boustani_master_2009}, an assumption that negates the possibility of collective oscillations.
The significant  progress provided by this alternative approach was initially limited by the use of abstract spiking neuron models not based on identifiable physiological parameters. 
Compared to the IK model, these models apply to only a limited range of neuron types and spiking patterns.
While numerous studies have extended mean-field theory to account for mechanisms such as spike-frequency adaptation \cite{gast_mean-field_2020}, synaptic plasticity \cite{taher_exact_2020,gast_mean-field_2021}, or gap junctions \cite{pietras_exact_2019,montbrio_exact_2020}, all of these studies were based on spiking neuron models written in terms of dimensionless variables and parameters.
As a result, these models do not provide a direct link between the model parameters and experimentally accessible quantities that characterize neural structure and function.

Here we presented a mean-field model that does provide such a link and may therefore be used to make experimentally testable predictions about the effect of the physiological properties of individual neurons on population dynamics.
For example, we can determine how changes in neural resting potentials, spike waveforms, or rate of spike-frequency adaptation can be expected to change population responses.
These predictions might be tested through direct experimental manipulation, or by studying how physiological properties naturally vary across cortical regions and layers, as in \cite{hodge2019conserved}, and relating these differences to cell population dynamics across regions.

To assess the feasibility of using mean-field techniques to study the population dynamics of physiologically relatable neural models, we analyzed the validity of the mean-field model under violations of three key assumptions that are required for the closed form derivation presented here.

We first examined the assumptions regarding the strength of spike-frequency adaptation in the spiking neural network.
Previous studies have demonstrated that spike-frequency adaptation has a critical impact on the emergence of synchronized states such as population bursting in networks of coupled excitatory neurons \cite{fuhrmann_spike_2002, gigante_diverse_2007, gast_mean-field_2020}; this is therefore an important element to include in mean-field population models.
To derive the equation for the mean-field dynamics of the average recovery variable $u$, we followed \cite{chen_exact_2022} in assuming that spike-frequency adaptation is weak in comparison to the magnitude of the recovery variable.
As expected, we found that the violation of this assumption decreases the agreement between the predicted mean-field dynamics and the actual dynamics of the spiking neural network.
However, our results suggest that the bifurcation structure of the spiking neural network is preserved in the mean-field model even when spike-frequency adaptation is strong.
Although the input intensity at which bifurcations occur is shifted relative to that of the spiking network, our mean-field model nonetheless captured the emergence of synchronized and bistable states observed in spiking neural networks.
We conclude that our mean-field model is a useful tool for analyzing population dynamics in the presence of spike-frequency adaptation.

We next examined the assumptions pertaining to the spike reset condition in the mean-field model.
As discussed in \cite{montbrio_macroscopic_2015,montbrio_exact_2020}, the derivation of the mean-field equations requires the assumption that IK neurons produce their spike when $v_i = v_p \rightarrow \infty$; upon spiking are reset to $v_i = v_0 \rightarrow -\infty$.
This assumption is particularly problematic when using biophysiological neuron models, as the membrane potentials reported of neurons fall within a relatively narrow range \cite{dayan_theoretical_2001}.
Furthermore, setting $v_0$ and $v_p$ to specific, finite values is needed for IK model neurons to reproduce the spiking dynamics of different biological neuron types \cite{izhikevich_simple_2003,izhikevich_large-scale_2008,humphries_dopamine-modulated_2009}.
Our results indicate that imposing realistic spike reset conditions mostly leads to an increase in the average firing rate of the spiking network model relative to the mean-field model.
We derived a rescaling of the background input to the network and showed that this adjustment is sufficient to correct for the increased firing rate introduced by realistic spike reset conditions.
This rescaled input can be used in the mean-field equations, leading to a significantly improved agreement between mean-field and spiking neural network dynamics.
We conclude that the mean-field model derived in this work can describe the different mean-field dynamics for spiking networks with distinct spike reset conditions.

Finally, we examined our assumption that neural spike thresholds follow a Lorentzian distribution.
In a biological system, values in the heavy tails of the Lorentzian will never be observed; the values of $v_\theta$ cannot exceed the peak potential $v_p$ or be lower than the reset potential $v_r$.
The spiking neural networks effectively exhibit a truncated Lorentzian distribution of spike thresholds with $v_r < v_\theta < v_p$, whereas the derivation of the mean-field model requires us to assume a full Lorentzian distribution.  
We found that the agreement between mean-field and spiking models depended on how strongly the distribution for $v_\theta$ was truncated in the spiking neural network; however, the difference between mean-field predictions and spiking neural network dynamics was small in comparison to the finite-size fluctuations of the latter. 
Furthermore, we found a good agreement between mean-field predictions and spiking neural network dynamics in all models examined in this work, once inaccuracies caused by spike frequency adaptation or narrow spike reset conditions were accounted for.
We conclude that the full Lorentzian approximation for the distribution of spike thresholds in the spiking neural network leads to accurate mean-field predictions of network dynamics.
This result implies that truncated Lorentzian distributions can be used to fit experimental measurements of spike thresholds in biological neural populations, and that our mean-filed model can then be used to analyze their population dynamics.
In \cite{gast_effects_2022} we used a similar approach to analyze the impact of spike threshold heterogeneities in different interneuron populations on the dynamics of mesoscopic brain circuits.
In this work we fitted the statistics of distinct spike threshold distributions measured in brain slices (see \cite{lau_impaired_2000,wang_anatomical_2004}) to truncated Lorentzian distributions and we analyzed the impact of spike threshold heterogeneities on the phase transitions of mesoscopic brain circuits. 

In conclusion, we have derived and analyzed a mean-field model of interacting heterogeneous spiking neurons.
Our detailed analysis of the mean-field model predictions provides a clear picture of the conditions under which the mean-field predictions can be expected to be an accurate representation of the dynamics of spiking networks.
As our mean-filed model was built upon IK neural models, it provides a degree of flexibility and biophysical detail that allows it to be applied to neural recordings in a wide range of brain regions and systems.

\section{Acknowledgements}

We would like to thank the Michael J. Fox Foundation for their support of RG via the Aligning Science Across Parkinson’s grant awarded to AK.

\bibliography{references}

\end{document}